\documentclass{llncs}
\usepackage{llncsdoc}
\usepackage{amsmath}
\usepackage{amsfonts}
\usepackage{amssymb}
\usepackage{tikz}
\usepackage{verbatim}
\usepackage{hyperref}
\usepackage{microtype}
\usepackage{booktabs}
\usepackage{algorithm}
\usepackage[noend]{algpseudocode}
\usepackage{float}
\floatstyle{plaintop}
\restylefloat{table}
\usepackage[tableposition=top]{caption}
\usepackage{subcaption}
\allowdisplaybreaks

\begin{document}
\newcommand{\ra}[1]{\renewcommand{\arraystretch}{#1}}

\title{Socio-Spatial Pareto Frontiers of Twitter Networks}

\author{Brandon Oselio \and Alex Kulesza \and Alfred Hero}

\institute{Department of Electrical Engineering and Computer Science, \\
University of Michigan, Ann Arbor, MI, USA \\
\email{ \{boselio, kulesza, hero\}@umich.edu}
\thanks{This work was partially supported by ARO under grant \#W911NF-12-1-0443. We are grateful to Qiaozhu Mei who provided the Twitter data stream through his API gardenhose level access.}}

\maketitle

\begin{abstract}
Social media provides a rich source of networked data. This data is represented by a set of nodes and a set of relations (edges). It is often possible to obtain or infer multiple types of relations from the same set of nodes, such as observed friend connections, inferred links via semantic comparison, or relations based off of geographic proximity. These edge sets can be represented by one multi-layer network. In this paper we review a method to perform community detection of multi-layer networks, and illustrate its use as a visualization tool for analyzing different community partitions. The algorithm is illustrated on a dataset from Twitter, specifically regarding the National Football League (NFL). 
\end{abstract}

\section{Introduction}
\label{sec:Intro}
Social networks increasingly comprise multiple types of connectivity information. The set of users form nodes, while the connectivity information form sets of edges. This information can either be observed directly from the data --- \textit{relational} edge sets --- or inferred using ancillary data that describes users --- \textit{behavioral} edge sets \cite{OsKuHe14}. A multi-layer network \cite{KiArBa14,OsKuHe14,MaRo13-2} is a framework that allows for nodes with multiple edge sets. Multi-layer networks place each type of connectivity in its own layer; the goal is to then analyze this structure.


In performing tasks like community detection on multi-layer networks, we seek a flexible method that allows for each layer to contribute to the  overall community structure at varying levels of strength. One method to do this is to incrementally combine single layer solutions while visualizing each community partition.

Prior work has analyzed two-community networks using Pareto frontiers \cite{OsKuHe14}.  In this paper we generalize those methods to handle complex networks with many communities, applying the resulting algorithm to socio-spatial Twitter networks focusing on the 2013 NFL playoffs.


\section{Methods}
\label{sec:Methods}
\subsection{Pareto Optimality}
The development of the algorithm will follow \cite{OsKuHe14}; it is briefly described here. First, the concept of Pareto optimality must be introduced. Pareto optimality stems from multi-objective optimization, also known as vector optimization. This field of research deals with problems where the goal is to optimize multiple objective functions simultaneously. 

Let $f_i: X  \rightarrow  \mathbb R$, with $X$ being our solution space, and consider the following minimization problem:

\begin{equation}
\min_x [f_1(x), \ldots,  f_k(x)].
\end{equation} 

The key definition is dominating and non-dominating points:

\begin{definition}
Let $x_1$ and $x_2$ be in the solution space $X$. We say that $x_1$ dominates $x_2$ if for all $i=1,\ldots,k$, $f_i(x_1) \le f_i(x_2)$, and at least for one index $j$, $f_j(x_1) < f_j(x_2)$. A solution $y$ in the solution space that is not dominated by any other point in the solution space is said to be non-dominated.
\end{definition}

The \textbf{Pareto front} of solutions is the set of solutions which are non-dominated. So, if a solution is on the Pareto front, it's not possible to achieve a lower value for a particular objective function without increasing the value of at least one other. The Pareto front defines a type of optimal set for vector optimization problems. 

\subsection{Pareto Front Algorithm}

Many single-layer community detection algorithms are objective based.  We can adapt these techniques to multi-layer networks by defining an objective function for each layer, and then exploring the Pareto front of community partitions.  

A multi-layer network $G = (\mathcal V, \mathcal E)$ consists of vertices $\mathcal V = \{ v_1, \ldots, v_p \}$, common to all layers, and edges $\mathcal E = (\mathcal E_1, \ldots, \mathcal E_L)$ in $L$ layers, where $\mathcal E_l$ is the edge set for layer $l$, and $\mathcal E_l = \{e^l_{v_i v_j}; \quad v_i, v_j \in V \}$.  Each edge is undirected. Further, a series of adjacency matrices are defined, one for each layer, where we have:

\begin{equation}
[[A^l]]_{ij} = e^l_{v_i v_j}.
\end{equation}

These adjacency matrices are important in order to evaluate the objective function for each layer. The goal is to find a community partition $C$ which divides the nodes into $k$ communities $B_1, \ldots B_k$, where $B_i \subseteq \mathcal V$. In this paper, RatioCut \cite{Lu07} is used as the objective function for each layer, defined as:

\begin{align}
  f_l(C) &= \frac{1}{2} \sum_{i = 1}^k\frac{\mathrm{cut}_l(B_i, \bar B_i)}{|B_i|} &
  \mathrm{cut}_l(B_i, \bar B_i) &= \sum_{i \in B_i, j \notin B_i} [A^l]_{ij}
\end{align}

In describing the algorithm, we specialize to $L=2$ layers. First we perform single-layer network community detection on both layers, resulting in two community partitions $C_1$ and $C_2$. Note that normalized spectral clustering is a relaxation of the problem of minimizing RatioCut, which we perform on each layer, and obtain $C_1$ and $C_2$. We assume that these solutions are optimal in their respective objectives. Thus, $C_1$ and $C_2$ both lie on the Pareto front; in order to find other points on an approximate Pareto front between the given community partitions, a node swapping technique is used, similar to the KL algorithm \cite{KeLi70}. Starting at one solution $C_1$ and ending at solution $C_2$, one node $i$ changes its community membership from $C_1[i]$ to the membership $C_2[i]$ at each step of the algorithm. In order to determine which nodes to swap, the RatioCut is calculated for each possibility. This is continued until the ending solution $C_2$ is reached. Finally, all the traversed partitions are filtered to find non-dominated solutions; these solutions form an approximate Pareto front. Algorithm \ref{alg:paretoswap} describes the traversal in more detail. This algorithm improves on  \cite{OsKuHe14} in that it allows for unequal communities in the resulting partition and it is effective for more than two communities, which is useful when applying to real social network datasets.

\begin{algorithm}
\caption{Pareto Frontier Algorithm}\label{alg:euclid}
\begin{algorithmic}[1]
\Procedure{PossibleFrontierPoints}{$A_1,A_2, C_1, C_2$}
\State $C_{cur}\gets C_1$
\State $t \gets 0$
\State Cost $\gets \infty$
\While{$C_{cur}\not=C_2$}
	\For{$i$ where $C_{cur}[i] \not=C_2[i]$} 
	\State $C_{cur}[i] \gets C_2[i]$
	\State Cost$[i] \gets \text{RatioCut} (A_2, C_{cur})$
	\State $C_{cur}[i] \gets C_1[i]$
	\EndFor
	\State $i^* \gets \min_i $ Cost$[i]$
	\State $C_{cur}[i^*] \gets C_2[i^*]$
	\State Memberships$[t] \gets C_{cur}$
	\State Cuts$[t] \gets (\text{Ratio-Cut} (A_1, C_{cur}), \text{Ratio-Cut} (A_2, C_{cur}))$
	\State $t \gets t + 1$
\EndWhile
\State \textbf{return} Memberships, Cuts
\EndProcedure
\end{algorithmic}
\label{alg:paretoswap}
\end{algorithm}

\section{Twitter Dataset}
\label{sec:Twitter}
Data to create a multi-layer network was obtained from the Twitter stream API at gardenhose level access during January of 2013. Tweets were filtered based on the availability of geolocation data. This geolocation information allowed for the creation of the first layer of the multi-layer network. For every pair of users $i$ and $j$, they were connected ($A_{ij} = A_{ji} = 1$) if the users were closer than a certain distance threshold $\delta$. The $\delta$ parameter changed based on the density of users and size of area that was being observed. This layer is called the coordinate network layer.


The second layer that is created utilizes hashtags to connect users. Hashtags are any words beginning with a \# sign. In this layer, a user $i$ and $j$ are said to be connected if they use the same hashtag from a specified set of hashtags over the one month period. In order to focus on a smaller set of users, specific hashtags were chosen that applied to an event or set of events that were occurring in this period; in this case the events were the National Football League (NFL) playoffs. The dataset was created by first filtering on four of the most popular pertinent hashtags in the three month time period: \#Ravens, \#49ers, \#Falcons, and \#Patriots. These correspond to the four NFL football teams that reached the end of the NFL playoffs for that year. A two-layer network consisting of the hashtag network layer and coordinate network layer ($\delta=50$) is analyzed. The resulting dataset contains 3456 nodes (Twitter users).



\section{Results}
\label{sec:Results}

\begin{figure}[ht]
  \centering
  \begin{subfigure}[t]{0.47\textwidth}
        \centering
        \includegraphics[width=\textwidth]{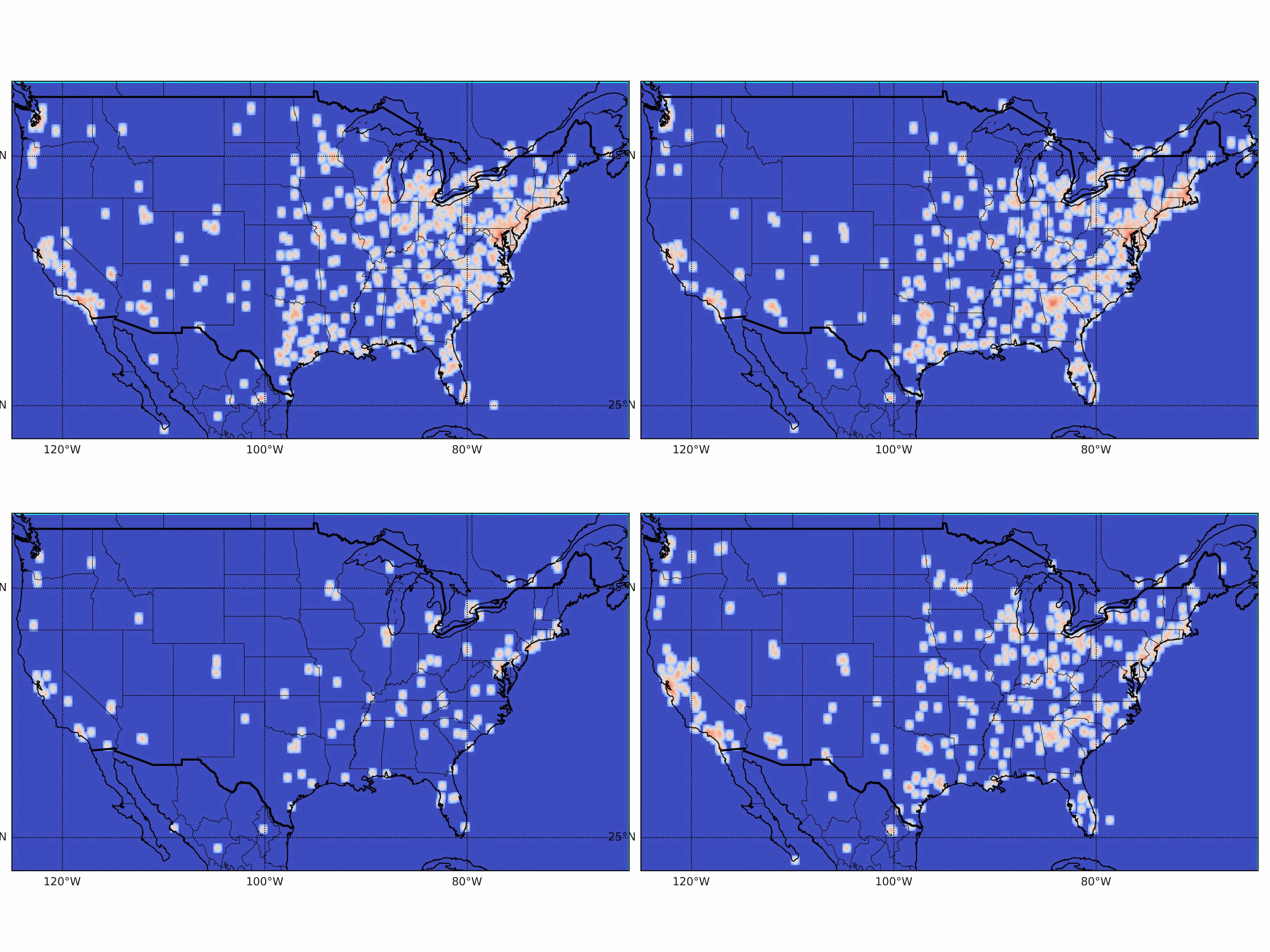}
        \caption{Hashtag Network Layer }
    \end{subfigure}
    \hfill
    \begin{subfigure}[t]{0.47\textwidth}
        \centering
       \includegraphics[width=\textwidth]{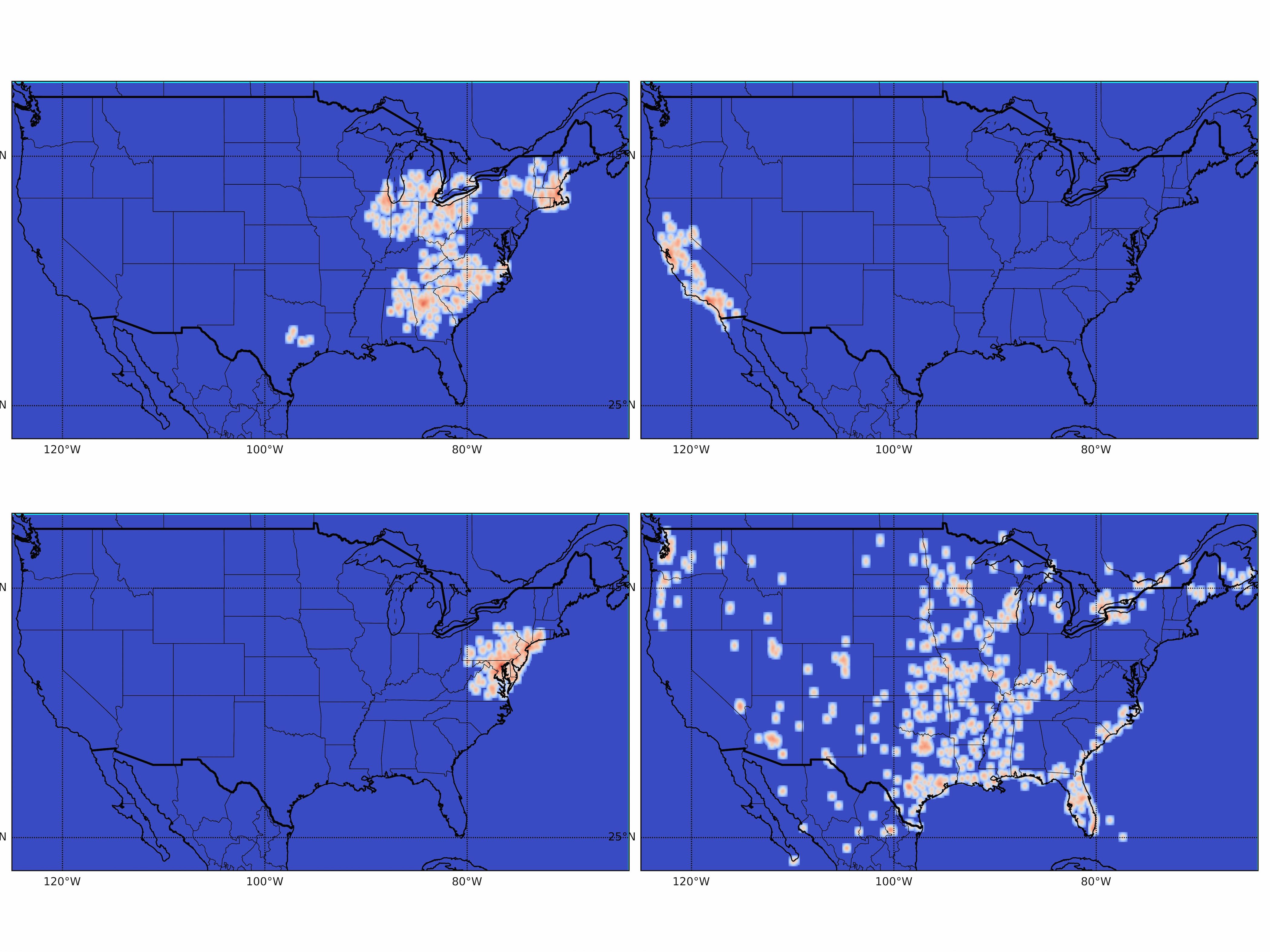}
        \caption{Coordinate Network Layer}
    \end{subfigure}
 
  \caption{Density Plot by Community. For the hashtag network layer, the communities correspond to the numbers in Table~\ref{table:football} going from left to right and subsequently from top to bottom. Communities based on the discussion of NFL teams are less localized than the communities based on geographic proximity.}
   \label{fig:Sports_T}
\end{figure}

We first perform some single-layer community detection. The partition resulting from spectral clustering on the hashtag network does a good job at stratifying the popular hashtags into communities, as seen in Table~\ref{table:football}. Community 1 is mostly the \#Ravens hashtag, while community 4 is the \#49ers. Community 2 sees the \#Patriots and \#Falcons hashtags grouped together, while community 3 is a mixture of all four. Figure~\ref{fig:Sports_T} shows a false color map of the densities of users in each community. It is surprising that while there is strong community structure in this network, it is less correlated with geography than one might expect.

\begin{table}
\ra{1.1}
\centering
\begin{tabular}{ccccc}
\toprule
		& Community 1 &  Community 2 & Community 3 & Community 4\\
\midrule

\#Ravens   & 1232 	& 0 			& 170 	& 0\\
\#49ers & 57 & 0 & 155 & 762\\
\#Patriots & 45 & 291 & 29 & 10\\
\#Falcons & 49 & 273 & 29 & 7\\
\bottomrule
\end{tabular}
\caption{Hashtags per Community for Hashtag Network Layer Solution}
\label{table:football}
\end{table}

\begin{figure}[ht]
  \centering
  \includegraphics[width = 0.60\textwidth]{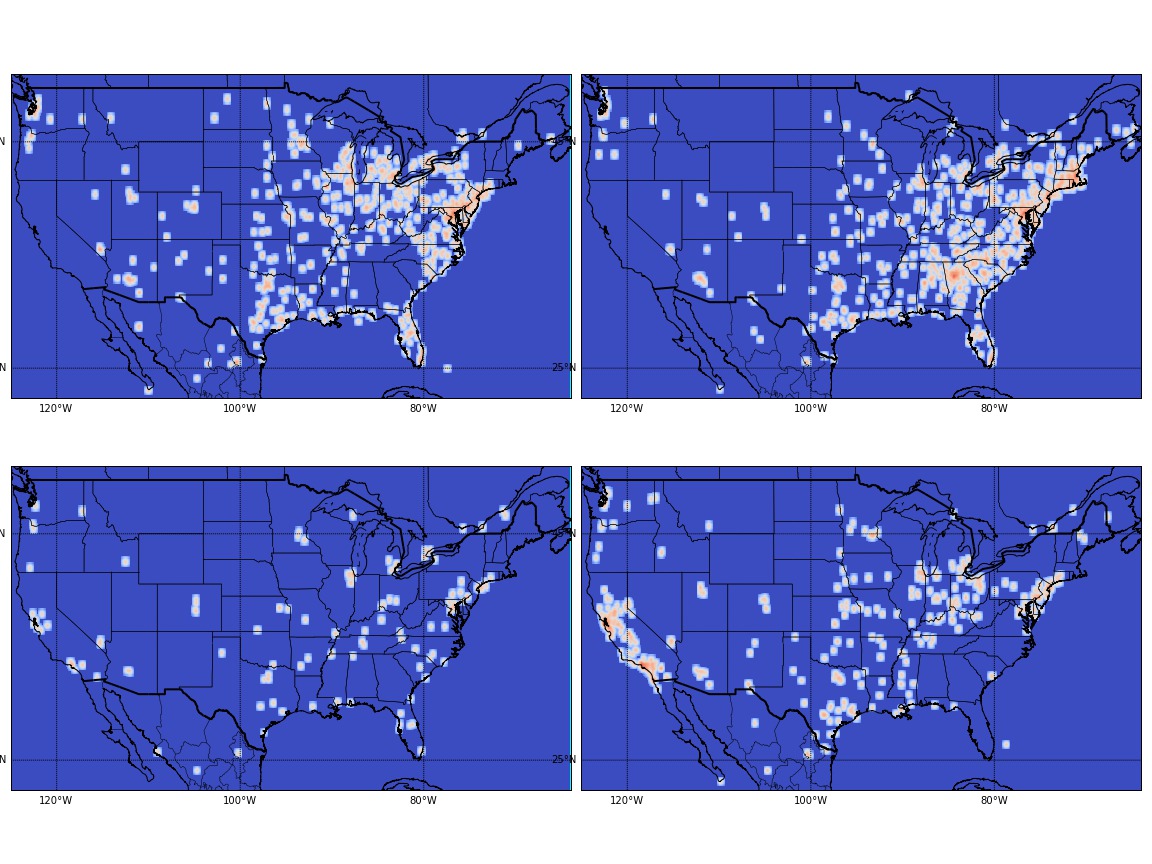}
  \caption{Density Plot for Pareto Combined Community. This community partition retains attributes from both layers, while still giving a visual sense of the overall community structure. The communities in the upper left and lower right have become more concentrated about east coast and west coast, respectively. Further, the community in the upper right shows high concentration in Atlanta up to both the Maryland and Massachusetts area.}
  \label{fig:Sports_P}
\end{figure}

As expected, the coordinate network layer partitions according to high population density. Specifically, it clusters the San Francisco and LA area together, the Maryland area by itself, and the Atlanta and Boston area together. The last community seems to be a catch-all for everywhere else, i.e., those places with less density.

Using the described algorithm, a Pareto solution is found for the multi-layer network; the community partition is shown in Figure~\ref{fig:Sports_P}. The communities are more geographically localized when compared with the mention network layer solution, while still visually resembling its structure. For instance, the last community picks out the San Francisco/LA area in a single community, which the original mention network did not. Further, the second community groups the Atlanta area with the Massachusetts area, though not as well as the coordinate network layer. The Pareto community partition, however, still contains some of the interesting patterns of the hashtag network layer and is not completely given to geographic localization.



\section{Related Work and Conclusion}
\label{sec:Related}
Work on multi-layer networks continues to increase; a more comprehensive overview of the techniques and theoretical background for the multi-layer structure can be found in \cite{KiArBa14}. Other methods have been proposed to utilize multiple types of data on nodes, including a mean approximation \cite{TaWaLi12}.

\cite{Fo10} provides an overview of the types of algorithms used for community detection. For multi-layer networks, the extension of modularity-type algorithms have been proposed \cite{MuRiMa13}. There have also been papers that analyzes single layer network community partitions when attempting to understand the multi-layer structure, as we are doing in this paper. \cite{BaFaMa11} compares single-layer communities via normalized mutual information (NMI), although does not try to combine the solutions in any way.  The particular single-layer algorithm used in this paper, normalized spectral clustering, has been well studied \cite{Lu07}. The Pareto front algorithm is an extension of the one found in \cite{OsKuHe14}.

This paper revisited and expanded upon the Pareto frontier algorithm which was detailed in \cite{OsKuHe14}. Extending the algorithm to multiple communities, the main purpose of this approach was to evaluate community partitions for varying levels of involvement by the layers of the network. The algorithm was applied to data from Twitter. The Pareto front method is useful to visualize community partitions for multi-layer datasets; this type of visualization is key to understanding how the layers are similar to each other, and how their community structures can interact. For future work, an explanation of how well we can approximate the Pareto front, as well as quantitative measurements on the similarity of community structure between layers would be helpful in the analysis of multi-layer networks.

\bibliographystyle{splncs}
\bibliography{SBP15}

\end{document}